# Catalogue of observed tangents to the spiral arms in the Milky Way galaxy

Jacques P. Vallée

National Research Council Canada, National Science Infrastructure portfolio, Herzberg Astrophysics, 5071 West Saanich Road, Victoria, B.C., Canada V9E 2E7

**Abstract.** From the Sun's location in the Galactic disk, one can use different arm tracers (CO, HI, thermal or ionized or relativistic electrons, masers, cold and hot dust, etc) to locate a tangent to each spiral arm in the disk of the Milky Way.We present a master Catalogue of the astronomically observed tangents to the Galaxy's spiral arms, using different arm tracers from the literature. Some arm tracers can have slightly divergent results from several papers, so a mean value is taken – see Appendix for CO, HII, and masers. The Catalogue of means currently consists of 63 mean tracer entries, spread over many arms (Carina, Crux-Centaurus, Norma, Perseus origin, near 3-kpc, Scutum, Sagittarius) , stemming from 107 original arm tracer entries.

Additionally, we updated and revised a previous statistical analysis of the angular offset and linear separation from the mid-arm, for each different mean arm tracer. Given enough arm tracers, and summing and averaging over all four spiral arms, one could determine if arm tracers have separate and parallel lanes in the Milky Way. This statistical analysis allows a cross-cut of a galactic spiral arm to be made, confirming a recent discovery of a linear separation between arm tracers. Here, from the mid-arm's CO to the inner edge's hot dust, the arm halfwidth is about 340 pc; doubling would yield a full arm width of 680 pc. We briefly compare these observations with the predictions of many spiral arm theories, notably the density-wave theory.

## 1. Introduction

Our whole Galaxy is composed of a disk orbiting around a nuclear bulge, all happening within a dark halo. At a distance between 7.5 and 8.5 kpc from the Galactic Center, the Sun is immersed inside the galactic disk, composed mainly of stars with the addition of dust (warm and hot), gas (molecular and atomic), ionized gas and thermal electrons, cosmic rays and relativistic electrons, and magnetic fields. For a review of some physical properties (density, temperature, etc) of the gas phases and the magnetic fields in the interstellar medium, see Heiles & Haverkorn (2012). Near the Galactic Center, there are two coexisting bar-like structures pointed between $14^{o}$ to $45^{o}$ with respect to the line of sight of the Sun to the Galactic Center, a stellar bulge, and the usual black hole at the center (Green et al 2011 – fig.5). The majority of published papers favors a 4-arm spiral model in the galactic disk. There is still a bit of controversy left, as one-sixth of all recent publications favor a 2-arm spiral structure (see Vallée 2014b).

Here our interest lies in the disk. Massive stars, and a large number of less massive star, molecular gas and dust are preferentially located in spiral arms. In the Milky Way, a spiral arm has a mean pitch angle near $12^o$, although estimates vary slightly. Previous sketches of the Milky Way galaxy showed stellar arms arranged in a spiral form, without much details across the arms. Recently, more details have become available.

The Sun is located in the interarm region (**Figure 1)**, between the Perseus arm and the Sagittarius arm. There is a local “armlet” (not shown here) of length about 2.2 kpc at a galactic longitude near $77^o$ (e.g., section 9 in Vallée 2011). There is no ‘molecular ring’ near the Galactic Center, but a molecular region encompassing the origins of the four spiral arms (Section 3.2 in Vallée 2008; Dobbs & Burkert 2012). The “Perseus origin arm” near the Galactic Center (Figure 1) is at times referred to as the “far 3-kpc arm” (e.g., Dame & Thaddeus 2008 and 2011, their $-23^o$ arm at $337^o$; table 3 in García et al 2014). The “Crux” arms is often called the “Centaurus” arm. The “Cygnus” arm is often called the “outer” arm. The “Cygnus + I’ arm is the continuation of the Scutum arm (Dame & Thaddeus – Fig.4). In Figure 1 a single bar is sketched (but no bulge), since the position angle of both bars are still loosely defined or explained (Martinez-Valpuesta 2013, theirFig.1; Romero-Gómez et al 2013, their fig.1). We know from the measured line-of-sight velocities at longitude $l=0^o$ that the near 3-kpc arm is expanding at -50 km/s (close to the origin of the Norma arm), and that the far 3-kpc arm is expanding at +50 km/s (close to the start of the Sagittarius arm). The author is not aware of any similar velocity measurement close to the start of the ‘Perseus origin’ arm, nor of the start of the Scutum arm.

There have been numerous published results for the spiral arms of the Milky Way. With the Sun in the Milky Way disk, the majority of published observational results have focused on small, different parts of our Galaxy, using different arm tracers, listing the arm shape (logarithmic or wavy or polynomial), the arm pitch angle (from the circular tangent), the number of arms (2 or 4 or 6), and the interarm separation through the Sun (around 3 kpc). Given conflicting results over time, statistics on arm parameters were done for the period 1980-onward, catalogued in blocks of 15 to 20 each (for the blocks, see Vallée 1995, 2002, 2005, 2008, 2013, 2014a, 2014b). A composite map of all these small-scale views could approximate the mid-scale view – see Figure 1.

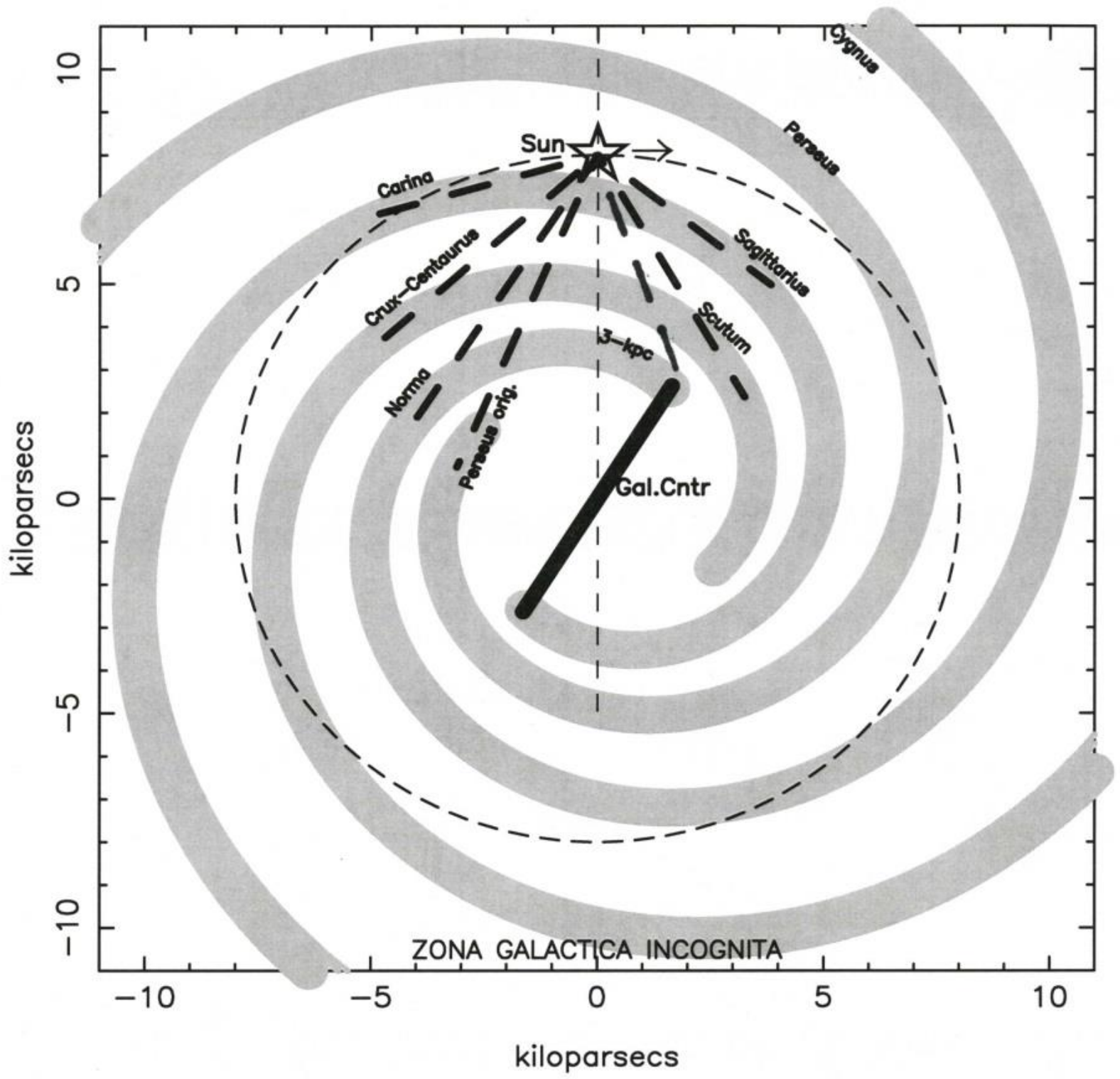

**Figure 1.** Old galactography. The sun is shown (open star) at 8 kpc from the Galactic Center, with the clockwise galactic rotation (arrow). Most stars in the Milky Way galaxy are located along 4 spiral arms (gray). Other component such as clouds, gas, atoms, dust, cosmic rays are also in there. Approximate arm tangents as seen from the sun are shown (thick black dashes). The sun's orbit is shown here as a circle (thin dashes). A rough position for the Galactic bar is shown. Not much is known below the Galactic Center, termed 'Zona Galactica Incognita' a.k.a. 'unknown galactic area' following Vallée (2002, his figure 2). The Galactic Center is at (0, 0) and the distance scale at bottom and to the left follows common conventions.

## 2. Catalogue of Galactic arm tangents in the Milky Way since 1980

Differing ratios of components (star, gas, dust, cosmic-ray, magnetic field) can be found along an arm; some arm "segments" may have more stars while others further on may have more molecular gas. Thus the fixed view from the Sun, along the galactic disk, can miss an arm segment if using a single filter (only CO data in one segment, or only stellar counts in another segment). If we could 'move' the Sun at different positions (segments) along the same arm, we may see different local views (differing component ratios) in optical (star), near infrared (dust), or radio wavelengths (relativistic or thermal electrons, molecular gas). It is thus necessary to use many tracers when talking about a single arm.

Small tables of arm tangents have been published before. Table 1 in Englmaier & Gerhard (1999) had 32 entries of tracer tangents, but none for the Carina arm near 284$^{o}$, their 'inner Galaxy' column mixed together the Scutum arm and the inner 3-kpc arm , and their far '3-kpc' arm near - 21$^{o}$ is now called the 'Perseus origin' arm. Table 2 in Vallée (2008) had 39 tracer entries, Table 1 in Vallée (2012) had 12 tracer entries, while Table 3 in Vallée (2014a) had 43 tracer entries.

Here we embarked on an extensive literature review, to glean more arm tangents from different tracers. We recorded the complete published reference of each tracer (journal, table, figure, section), and thus were able to delete or replace erroneous or misplaced references. Also, newer instrumentation allows the detection of the arm tangents as inferred from the recent 6.7 GHz methanol maser or water maser data. The master catalogue of published galactic longitudes of arm tangents is given in **Table 1**. In Table 1, an arm is defined and labeled in only one Galactic Quadrant (I or IV); its continuation in another quadrant has another name, and we listed 63 different entries of the means of different tracer tangents (one specific tracer entry per arm), about half more than in Vallée (2014a). As much as 17 of these 63 different entries are themselves a mean of two or more published values. In addition, in **Appendix A**, we provided 38 individual CO entries (Table 3), 13 individual HII entries (Table 4), and 10 individual maser entries (Table 5). Thus, all in all, we catalogued 107 different tracer entries over these four tables (63 -17 +38 +13 +10). For the Centaurux-Crux arm, the CO arm tangent was originally given as 310$^{o}$ by Bronfman et al (1988 – fig.7), then appeared as a typo as 300$^{o}$ in Bronfman (1992 – his Fig.6), and was later given as 309$^{o}$ in Garcia et al (2014 – their Fig. 14); in Appendix A we entered only the original value of 310$^{o}$ from the first publication in 1988.

We provide here a short CO analysis of the published results. The Columbia CO Surveys made at low angular resolution (~ 8 arcmin) of the low excitation (J=1-0), low temperature (~10 K), $^{12}$CO tracer (emitting at115 GHz), as integrated over a velocity range associated with a spiral arm, is useful to indicate the middle of the arm. For the Carina arm, the CO tracer is given in the literature with a longitude varying from 280$^{o}$ to 283$^{o}$, depending on the different mathematical analyses done. The different CO analyses gave slightly different output values from roughly the same observed data, especially when using different averaging bin sizes in galactic longitudes (asymmetric amplitude with galactic longitude), or using different input data cutoff (63% removal, representing the disk emission with longitude – see Fig. 14 in García et al 2014), or using a simple or complex disk emission model (Section 2 in Bronfman 1992). Of all the CO tracers used as arm tangents, the longitude-velocity analysis of García et al (2014) gave the largest longitude values for the Crux-Centaurus arm, Norma arm, and Perseus Origin (3-kpc) arm (their Table 3). They noted that their arms were first defined and delineated in the longitude-velocity diagram (their Fig. 9), and later transposed and averaged in the longitude-distance diagram. Such an analysis is bound to use somewhat arbitrary boundaries in the longitude-velocity diagram (removing some reassigned arm clouds), and the arm tangent obtained later in the other diagram with the transposed clouds could have a bigger error bar. The low resolution (8') Columbia CO survey (linked to the mid-arm) does differ as it should from the high resolution (50") Stoney Brook survey of warm CO cores (linked to new star forming regions - Solomon et al 1985).

**Table 1 – Catalogue of published different spiral arm tracers (since 1980), with only one mean tracer value for each arm**

| Arm Name | Chemical tracer | Gal. longit. of arm tangent[1] | Ang. dist.[2] to $^{12}$CO | Linear separation inside arm[3] from $^{12}$CO | References[4] |
|---|---|---|---|---|---|
| Carina | $^{12}$CO at 8’ | 282° | 0° | 0 pc, at 5 kpc[3] | Bronfman et al (2000b – table 2); see Table 3 |
| | Thermal electron | 283° | 1° | 87 pc | Taylor & Cordes (1993 – fig.4) |
| | HII complex | 284° | 2° | 174 pc | Russeil (2003 – table 6); see Table 4 |
| | Dust 240μm | 284° | 2° | 174 pc | Drimmel (2000 –fig.1) |
| | Dust 60μm | 285° | 3° | 262 pc | Bloemen et al (1990 – fig.5) |
| | FIR [CII] & [NII] | 287° | 5° | 435 pc | Steiman-Cameron et al (2010 – sect. 2.1) |
| Crux- | $^{12}$CO at 8’ | 309° | 0° | 0 pc, at 6 kpc[3] | Bronfman et al (2000b – table 2); see Table 3 |
| Cen- | Thermal electron | 309° | 0° | 0 pc | Taylor & Cordes (1993 – fig.4) |
| taurus | HII complex | 309° | 0° | 0 pc | Russeil (2003 – table 6); see Table 4 |
| | FIR [CII] & [NII] | 309° | 0° | 0 pc | Steiman-Cameron et al (2010 – sect. 2.1) |
| | HI atom | 310° | 1° | 105 pc | Englmaier & Gerhard (1999- table 1) |
| | $^{26}$Al | 310° | 1° | 105 pc | Chen et al (1996- fig.1) |
| | Sync. 408 MHz | 310° | 1° | 105 pc | Beuermann et al (1985 – fig.1) |
| | Dust 240μm | 311° | 2° | 209 pc | Drimmell (2000 – fig. 1) |
| | Dust 60μm | 311° | 2° | 209 pc | Bloemen et al (1990 – fig.5) |
| | Dust 870μm | 311° | 2° | 209 pc | Beuther et al (2012 – fig.2; another peak at 305°) |
| Norma | HII complex | 325° | -3° | -366 pc | Downes et al (1980 – fig.4); see Table 4 |
| | $^{26}$Al | 325° | -3° | -366 pc | Chen et al (1996 – fig.1) |
| | $^{12}$CO at 8’ | 328° | 0° | 0 pc, at 7 kpc[3] | Bronfman et al (2000b – table 2); see Table 3 |
| | Thermal electron | 328° | 0° | 0 pc | Taylor & Cordes (1993 – fig.4) |
| | HI atom | 328° | 0° | 0 pc | Englmaier & Gerhard (1999 – table 1) |
| | Sync. 408 MHz | 328° | 0° | 0 pc | Beuermann et al (1985 – fig.1) |
| | Dust 60μm | 329° | 1° | 122 pc | Bloemen et al (1990 – fig.5) |
| | Methanol masers | 331.5° | 3.5° | 427 pc | Caswell et al (2011 – sect. 4.6.2); see Table 5 |
| | Dust 240μm | 332° | 4° | 488 pc | Drimmell (2000 – fig. 1) |
| | Dust 2.4μm | 332° | 4° | 488 pc | Hayakawa et al (1981 – fig.2a) |
| | Dust 870μm | 332° | 4° | 488 pc | Beuther et al (2012 – fig.3) |
| Start | $^{12}$CO at 8’ | 337° | 0° | 0 pc, at 8 kpc[3] | Bronfman et al (2000b – table 2); see Table 3 |
| of Per- | FIR [CII] & [NII] | 338° | 1° | 140 pc | Steiman-Cameron et al (2010 – sect. 2.1) |
| seus | Dust 870μm | 338° | 1° | 140 pc | Beuther et al (2012 – fig.3) |
| | Methanol masers | 338° | 1° | 140 pc | Green et al (2011 – sect. 3.3.1); see Table 5 |
| | Sync. 408 MHz | 339° | 2° | 279 pc | Beuermann et al (1985 – fig.1) |
| | Dust 2.4μm | 339° | 2° | 279 pc | Hayakawa et al (1981 – fig.2a) |
| | Dust 60μm | 340° | 3° | 419 pc | Bloemen et al (1990 – fig.5) |
| Near | $^{12}$CO at 8’ | 026° | 0° | 0 pc, at 6 kpc[3] | Cohen et al (1980 – fig.3); see Table 3 |
| 3-kpc | HII complex | 025° | 1° | 105 pc | Russeil (2003 – table 6); see Table 4 |
| Arm | Warm $^{12}$CO cores | 024° | 2° | 209 pc | Solomon et al (1985 – fig 1b); Bania (1980 – fig.7) |
| Scutum | $^{12}CO$ at 8’ | 033° | 0° | 0 pc, at 5 kpc(3) | Sanders et al (1985 – fig. 5b); see Table 3 |
| | HII complex | 032° | 1° | 87 pc | Russeil (2003 – Table 6); see Table 4 |
| | $^{26}Al$ | 032° | 1° | 87 pc | Chen et al (1996 – fig.1) |
| | $^{13}CO$ | 032° | 1° | 87 pc | Stark & Lee (2006 – fig.1, v= +95 km/s) |
| | Thermal electron | 032° | 1° | 87 pc | Taylor & Cordes (1993 – fig.4) |
| | Sync. 408 MHz | 032° | 1° | 87 pc | Beuermann et al (1985 – fig.1) |
| | Dust 870µm | 031° | 2° | 174 pc | Beuther et al (2012 – fig.3) |
| | Dust 240µm | 031° | 2° | 174 pc | Drimmell (2000 – fig. 1) |
| | Warm $^{12}CO$ cores | 030° | 3° | 262 pc | Solomon et al (1985 – fig. 1b) |
| | FIR [CII] & [NII] | 030° | 3° | 262 pc | Steiman-Cameron et al (2010 – sect 2.1) |
| | HI atom | 029° | 4° | 349 pc | Englmaier & Gerhard (1999 – table 1) |
| | Dust 2.4µm | 029° | 4° | 349 pc | Hayakawa et al (1981 – fig. 2a) |
| | Methanol masers | 028° | 5° | 435 pc | Green et al (2011 – sect. 3.3.1); see Table 5 |
| | Dust 60µm | 026° | 7° | 609 pc | Bloemen et al (1990 – fig.5) |
| | | | | | |
| Sagit-tarius | $^{12}CO$ at 8’ | 051° | 0° | 0 pc, at 4 kpc(3) | Cohen et al (1980 – fig.3); see Table 3 |
| | HII complex | 051° | 0° | 0 pc | Russeil et al (2007 – fig.4); see Table 4 |
| | $^{13}CO$ | 051° | 0° | 0 pc | Stark & Lee (2006 – fig.1, v= +60 km/s) |
| | HI atom | 050° | 1° | 70 pc | Englmaier & Gerhard (1999 – table 1) |
| | Dust 240µm | 050° | 1° | 70 pc | Drimmell (2000 – fig. 1) |
| | Methanol masers | 050° | 1° | 70 pc | Reid et al (2014 – fig.1); see Table 5 |
| | FIR [CII] & [NII] | 050° | 1° | 70 pc | Steiman-Cameron et al (2010 – sect. 2.1) |
| | Warm $^{12}CO$ cores | 049° | 2° | 140 pc | Solomon et al (1985 – fig. 1b) |
| | Dust 870µm | 049° | 2° | 140 pc | Beuther et al (2012 – fig.3) |
| | Thermal electron | 049° | 2° | 140 pc | Taylor & Cordes (1993 – fig.4) |
| | Sync. 408 MHz | 048° | 3° | 209 pc | Beuermann et al (1985 – fig. 1) |
| | $^{26}Al$ | 046° | 5° | 449 pc | Chen et al (1996 – fig.1) |

Notes:

(1): Galactic longitude of arm tangent, as observed by each tracer (not theoretically computed).

(2): Angular distance from arm center, being positive towards arm's inner edge (towards Galactic Center), and negative in other direction (galactic anti-center).

(3): Linear separation from the arm center ($^{12}CO$), after converting the angular separation at the arm distance from the sun (taken from Figure 1 here); we assume 8.0 kpc for the distance of the Sun to the Galactic Center.

(4): When there are more than 2 published reports for a given arm tracer at a given spiral arm, then a separate table is provided in the Appendix.

**Table 2 - The linear separation (S) from $^{12}CO$ of each arm tracer, in each spiral arm in the Milky Way[1]**

| Chemical Tracer | S in Carina arm (pc) | S in Crux arm (pc) | S in Norma arm (pc) | S in Start of Per-Seus arm (pc) | S in Near 3-kpc arm (pc) | S in Scutum arm (pc) | S in Sa-git-ta-rius arm (pc) | Mean sepa-ration (pc) | s.d.m.[2] (pc) |
|---|---|---|---|---|---|---|---|---|---|
| $^{12}CO$ at 8’ | 0 | 0 | 0 | 0 | 0 | 0 | 0 | 0 | -[3] |
| HII complex | 174 | 0 | -366 | - | 105 | 87 | 0 | 0 | 78 |
| $^{13}CO$ | - | - | - | - | - | 87 | 0 | 44 | 44 |
| Thermal electron | 87 | 0 | 0 | - | - | 87 | 140 | 63 | 27 |
| $^{26}Al$ | - | 105 | -366 | - | - | 87 | 449 | 69 | 167 |
| HI atom | - | 105 | 0 | - | - | 349 | 70 | 131 | 76 |
| Synch. 408 MHz | - | 105 | 0 | 279 | - | 87 | 209 | 136 | 93 |
| FIR [CII] & [NII] | 435 | 0 | - | 140 | - | 262 | 70 | 181 | 77 |
| Warm $^{12}CO$ cores | - | - | - | - | 209 | 262 | 140 | 204 | 35 |
| Cold dust 240µm | 174 | 209 | 488 | - | - | 174 | 70 | 223 | 70[4] |
| Cold dust 870µm | - | 209 | 488 | 140 | - | 174 | 140 | 230 | 46[4] |
| Methanol masers | - | - | 427 | 140 | - | 435 | 70 | 268 | 95 |
| Hot dust 60µm | 262 | 209 | 122 | 419 | - | 609 | - | 324 | 86[5] |
| Hot dust 2.4µm | - | - | 488 | 279 | - | 349 | - | 372 | 61[5] |

Notes:

(1): All data from Table 1 here.

(2): The s.d.m. in the last column is from the external scatter.

(3): There is a mean internal scatter of 42 pc, from the CO data in the Appendix (Table 3).

(4): Stats made on both cold dust tracers give a mean separation of 227 pc, with s.d.m. of 46 pc.

(5): Stats made on both hot dust tracers give a mean separation of 342 pc, with s.d.m. of 56 pc.

### 3. Arm tangent offset, for each tracer

Different arm tracers appear at slightly different galactic longitudes. To transfer angular to linear offsets, we used the canonical 4-arm logarithmic spiral model, with a Sun-Galactic Centre distance of 8.0 kpc, which is a mean of many recent measurements of this distance. The observed angular offset and computed linear separation of each individual arm tracer from the mid-arm are given, using the galactic longitude of the arm tangent for tracer X versus the galactic longitude of $^{12}$CO arm tangent at low angular resolution.

Angle-wise, it can readily be seen in Table 1 that the hot dust arm tangents (and the methanol masers) are always inward, closer to the direction of the Galactic Center. Given enough arm tracers, and summing and averaging over all four spiral arms, one could determine if arm tracers have separate and parallel lanes in the Milky Way.

In **Table 2** we pull together the statistics over several spiral arms in the Milky Way, indicating the mean separation for each arm tracer, from the mid-arm. A net separation can be found for some of the arm tracers, notably for $^{12}$CO (large beam), thermal electrons (galactic free electrons, from pulsar rotation measure and dispersion measure), relativistic electrons (408 MHz synchrotron), hot dust (60μm and 2.4μm together), and methanol masers (6.7 GHz). The $^{26}$Al and HII complex tracers have large error bars, with a mean consistent with zero offset.

Here we make a short analysis of published stellar counts in the near infrared. The Glimpse program from the Spitzer Space Telescope has generated claims about alternating major and minor spiral arms (broad peaks in Fig.1 in Benjamin 2009; Fig. 14 and Fig.15 in Churchwell et al 2009), proposing the disappearance of the Sagittarius arm in stellar counts at λ4.5μm; some authors have questioned their analysis. Steiman-Cameron (2010) and Steiman-Cameron et al (2010, Sect. 4.1) pointed out that the Sagittarius arm is observed in every major tracer, and that there is towards that galactic longitude an unusual high dust absorption, enhanced obscuration, and increased extinction (around the W51 complex). Not properly accounting for this extra dust could cause a decrease in star counts and the arm's disappearance. Also, the arm width of their major arm (Fig. 15 in Churchwell et al, 2009) is about 2.5 larger than that of their minor arm; this was questioned later. Vallée (2014a – Table 4) pointed out that the width of the Sagittarius arm is equal or nearly equal to that of the other arms, when using at least 4 different arm tracers.

### 4. Galactic arm cross-cut

#### 4.1 Observational results

A crosscut of a spiral arm (**Figure 2)** shows the approximate position of a typical arm tracer with respect to the mid-arm, for the Milky Way. Many arms were used for the cross-cut. The direction to the Galactic Center is to the right. It is interesting that in Fig.2 the hot dust and methanol masers are always to the right of the $^{12}$CO arm tangents, while the relativistic electron and free-free electron are always in between the mid-arm and the dust lane. There is no observed spiral arm tracer between the middle of the arm and the arm's outer edge, facing the outer Galaxy, except for stars (see Table 2).

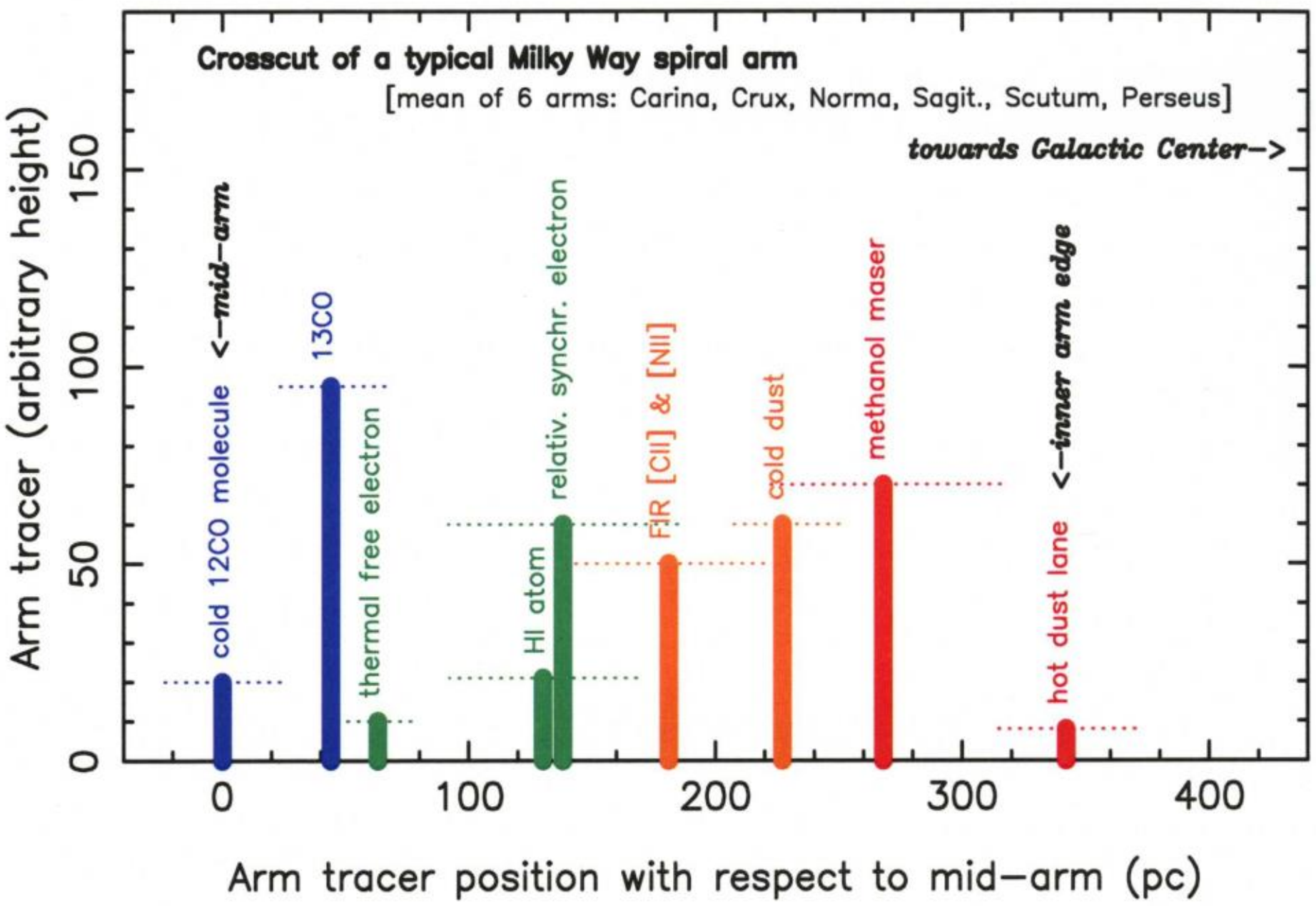

**Figure 2**. Average crosscut of various spiral arm in the Milky Way, showing where each different arm tracer appears. The sketch indicates the mean position of each spiral arm, as an offset from the mid-arm ($^{12}$CO tangent, at 0 pc) to the inner arm edge (hot dust tangent, near 340 pc). Vertical bars are arbitrary. The standard deviation of the mean (s.d.m.) is noted as an horizontal dotted line, for each arm tracer. Here the relativistic synchrotron is at 408 MHz. All data are from Table 2. For greater precision, the hot dust at 342 pc is an average over two wavelengths (2.4μ and 60μ) ; the cold dust at 227 pc is an average over two wavelengths (240μ and 870μ). For greater clarity, some tracers were omitted: the average HII complex at 0 pc, the $^{26}$Al at 69 pc, and the warm CO cores at 204 pc.

The observed half width of a typical arm (near 340 pc from mid-arm to the inner edge) can be doubled in order to get the full arm width, giving about 680 pc. The standard deviation of the mean (s.d.m.) is given for each arm tracer with the horizontal dotted lines. The mean sdm observed for the arm tracers shown (including $^{12}CO$) is near 80 pc (from **Table 2**), about only 12% of the full arm width.

The latest galactographic view of the Milky Way (**Figure 3)** shows the spiral arm lanes, using the data from Table 2 and Figure 2. On its way in a roughly circular orbit around the Galactic Center, the Sun would eventually reach the Perseus arm's hot dust lane first, and then later the next lanes (electrons; …; CO) before exiting on the other arm's side.

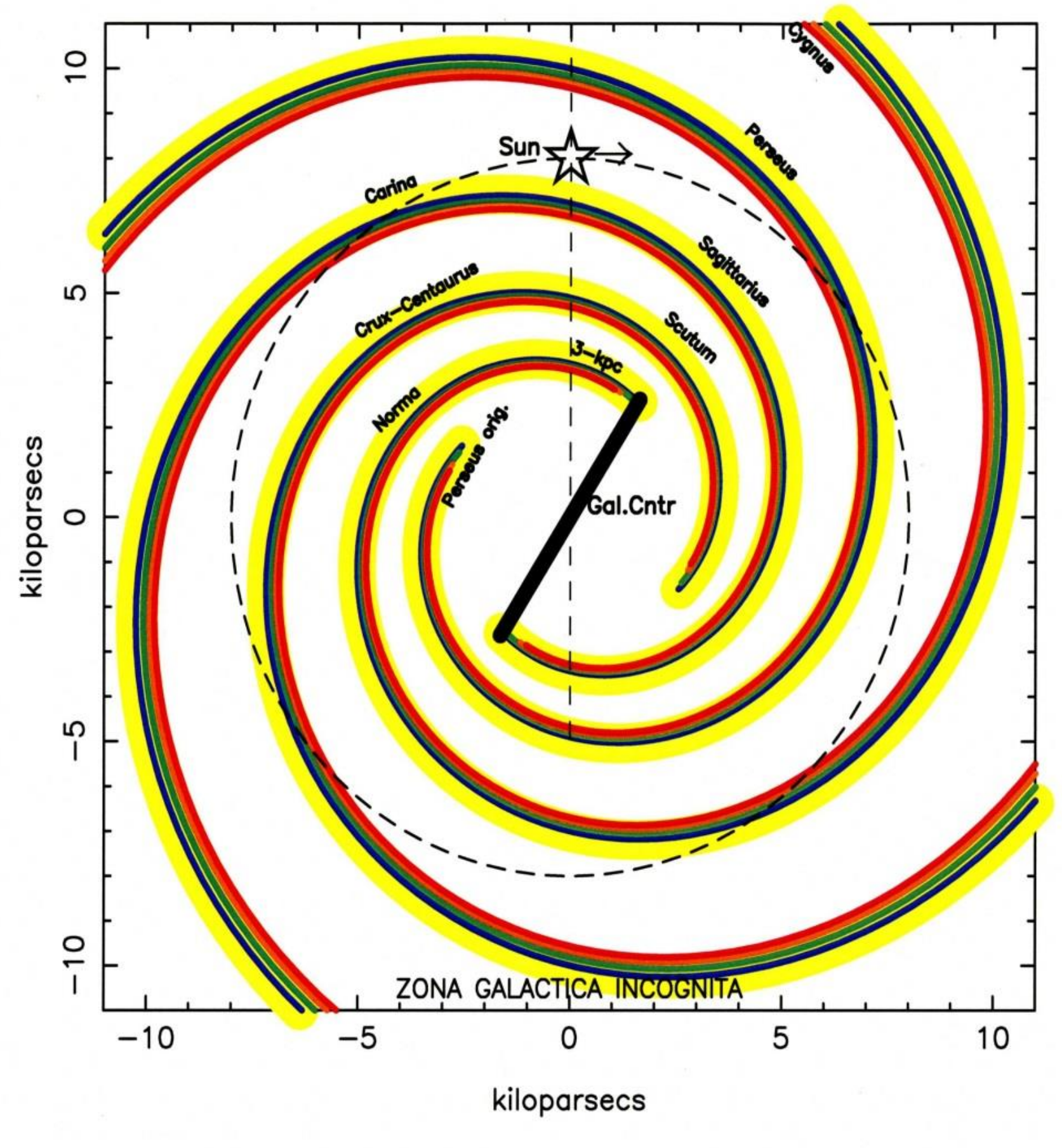

**Figure 3.** New galactography. Besides stars all over spiral arms, spiral arms harbor different components (tracers) at slightly at different (parallel) locations. Stars (yellow) are all over the arms. The $^{12}CO$ (and $H_2$) molecules peak near the middle of the arm (blue), while the hot dust (with masers and newborn stars) peak near the inner arm edge (red). Other intermediate tracers peak in between: thermal and relativistic electrons and HI atoms (green), and cold dust with FIR cooling lines (orange). Same color coding used as in Figure 2. Updated from Vallée (2014a).

**4.2 Comparisons with predictions from theories**.

Here we make an attempt at some quick interpretation in terms of current spiral arm theories, basically trying to relate our results with their predictions. In this brief assessment of theories of spiral arms, we compare a few observed arm parameters: linear separation of different arm tracers from $^{12}CO$ (table 2), no arm tracer to the left of $^{12}CO$ (figure 2), the number of arms (4) , average pitch angle ($12^o$), arm shape (logarithmic) and arm spacing (equal), to first order – see Vallée (2014a ).

The quasi-stationary density-wave theory can produce several arms. (i) The density wave theory predicts an orbital trajectory of the gas and stars going through spiral arms, and a linear separation between arm tracers as occasioned by a shocked gas and dust entering the arm from the inner edge, then some of it coalescing to form stars there (dust lane), progressing through the arm and exiting on the arm's outer edge (Roberts, 1975 – fig.3); this is similar to our orbit in Figure 1 here. (ii) In the density wave, the width of the dust lane (shocked lane) was predicted at 50 pc (e.g., Dobbs & Baba 2014, their sect. 3.5); this width is similar to our observed dust lane in Figure 2 here (with an error bar of 70 pc). (iii) In the density wave, the gas density maximum (at mid-arm) is separated from the shock lane by about 12 Myrs (Roberts, 1975 – Fig.2), or 306 pc for a gas velocity difference of 25 km/s (Dobbs & Baba 2014 – fig. 17b); this separation is similar to our observed arm's halfwidth of 360 pc (Table 2 here). Roberts (1975 – Fig.4) used a pitch angle near $10^o$, similar to that of the Milky Way. In the density-wave with 4 arms, the gas density maximum (potential minimum, broad CO at mid-arm) is also separated from the shock lane (hot dust) in the model of Gittins and Clarke (2004 – fig.16), and their 'standard model' has a low pitch angle of $5^o$ (not the $12^o$ found in the Milky Way). (iv) The density wave theory proposed that the peak of all tracers are between the inner edge of an arm and the mid-arm (as seen here in Table 2), excepting the stars distributed all over the arm. (v) The density wave predicts that the peak of the HII complex is located near the mid-arm (Roberts 1975 – fig.4, Gittins & Clarke 2004 – fig.16 far from the Galactic Center), as seen in Table 2 here.

There is a plethora of theories to generate spiral arms, in addition to the density-wave theory. (vi) The tidal theory involving a recent passage of the Sagittarius galaxy could induce two strong arms (Dobbs & Baba (2014, sect. 5.3), not four as found in the Milky Way. Of course, each of two Magellanic Clouds could perhaps produce its own set of twin arms. (vii) The bar-driven theory could induce two strong arms (Dobbs & Baba 2014 – sect.2.3), not four as found in the Milky Way. Of course, each of two straddling bars could perhaps produce its own set of twin arms. (viii) The stochastic, self-propagating star formation theory could produce a very irregular, flocculent arm pattern (Dobbs & Baba – sect. 2.5), not the quasi-regular pattern seen in the Milky Way. (ix) The swing-amplification theory could produce numerous flocculent arms irregularly located (Dobbs & Baba – sect. 5.2), not quasi-regularly spaced as for the Milky Way. (x) The transient-recurrent dynamic spiral theory (Dobbs & Baba 2014 – Sect. 2.2) has arms breaking and reconnecting, and high pitch angle between $20^o$ and $40^o$ (Dobbs & Baba 2014 – fig.10), unlike the low $12^o$ as found generally in the Milky Way. (xi)The basic dynamo theory does not predict any spiral arm as defined here (with dust and stars), contrary to observations, although it can produce magnetic arms – for a discussion, see Vallée (2011 – Sect. 9.7). In the presence of gaseous spiral arms, non-linear galactic dynamos predict that the interarm magnetic pitch angle will be compressed to roughly align with the gaseous arm pitch angle (Chamandy et al 2014 – sect. 7). (xii) The basic MHD theory, subjected to a spiral potential, predicts different pitch angle of the magnetic field (along the gas velocity) and that of the spiral arm (Dobbs & Price – fig.13); this is not at odd with some magnetic field observations – for a discussion, see Vallée (2011 – Sect.9.7.3).

In summary, the arm observations so far may be best compared to the density waves (i-v), but many other proposed theories could be partially contributing to the creation or maintenance over time of spiral arms or spurs (vi-xii). How much from each is unknown.

## 5. Conclusion

Our place in the Milky Way galaxy (Figure1), inside the flat galactic disk, makes it difficult to see the different spiral arms, being one behind another. Strong dust absorption in the disk does reduce our view at optical wavelengths. Still, radio, infrared and optical telescopes, equipped with spectrometers and other instruments, have allowed a picture to emerge with 4 long spiral arms where most of the gas and stars are positioned and their radial velocity measured. We presented here an updated and revised Catalogue of the mean data on the different spiral arm tracers (Table 1) and their accumulations (Table 3, Table 4, Table 5), covering 105 tracer entries.

We thus confirm here a separation between arm tracers (Table 2, Fig. 2), as claimed earlier by Vallée (2014a) with fewer tracers. In the Milky Way, each spiral arm tracer can occupy a parallel lane inside an arm – there is roughly 340 pc separation between the mid-arm ($^{12}$CO tracer lane) and the inner arm edge toward the Galactic Center (hot dust lane). This provides us a fascinating view of the inside of a typical spiral arm in the Milky Way galaxy, despite the Sun's awkward position inside the flat galactic disk (Figure 3). Our interpretation in terms of the many proposed spiral arm theories is still sketchy (section 4.2), while currently favoring the hunt for the density waves.

**Acknowledgements.**

The figure production made use of the PGPLOT software at the NRC-nsip in Victoria. I thank a referee for insightful and helpful comments, which helped to correct and improve the clarity of the text.

## Appendix A.

Here are some of the published individual tracer values, when there is more than one for each tracer in each spiral arm.

**Table 3** lists numerous individual CO entries (not earlier than 1980).

**Table 4** provides individual HII complex entries (not earlier than 1980).

**Table 5** gives individual maser entries (not earlier than 1980).

**Table 3 - Published CO J=1-0 results at low angular resolution (since 1980)**

| Arm name | Tangent Longitude | Telescope HPBW and survey name | | Reference |
|---|---|---|---|---|
| Carina | 280$^o$ | 8.8' | Columbia | Alvarez et al (1990 – table 4) |
| | 280$^o$ | 8.8' | Columbia | Grabelsky et al (1987 – sect. 3.1.2) |
| | 281$^o$ | 8.8' | Columbia | Grabelsky et al (1988 – fig.4) |
| | 282$^o$ | 8.8' | Columbia | Bronfman et al (2000b – table 2) |
| | 283$^o$ | 8.8' | Columbia | Bronfman et al (2000a – Section 3.4) |
| | 281.2 ±1.3 | mean and r.m.s. | | s.d.m of 0.6$^o$, worth 51 pc at 5 kpc |
| Crux-Cen-Taurus | 308$^o$ | 8.8' | Columbia | Bronfman et al (2000a – sect. 3.4) |
| | 308$^o$ | 8.8' | Columbia | Bronfman (2008 – sect. 4) |
| | 309$^o$ | 8.4' | CfA | Dame & Thaddeus (2011 – fig.4) |
| | 309$^o$ | 8.8' | Columbia | Bronfman et al (2000b – table 2) |
| | 310$^o$ | 8.8' | Columbia | Alvarez et al (1990 – table 4) |
| | 310$^o$ | 8.8' | Columbia | Bronfman et al (1988 – fig. 7) |
| | 310$^o$ | 8.8' | Columbia | Bronfman et al (1989 – sect. 4) |
| | 310$^o$ | 8.8' | Columbia | García et al (2014 – table 3) |
| | 310$^o$ | 8.8' | Columbia | Grabelsky et al (1987 – sect. 3.1.2) |
| | 309.3 ±0.9 | mean and r.m.s. | | s.d.m of 0.3$^o$, worth 34 pc at 6 kpc |
| Norma | 328$^o$ | 8.8' | Columbia | Alvarez et al (1990 – table 4) |
| | 328$^o$ | 8.8' | Columbia | Bronfman et al (1988 – fig. 7) |
| | 328$^o$ | 8.8' | Columbia | Bronfman et al (1989 – sect. 4) |
| | 328$^o$ | 8.8' | Columbia | Bronfman (1992 – fig.6) |
| | 328$^o$ | 8.8' | Columbia | Bronfman et al (2000a – sect. 3.4) |
| | 328$^o$ | 8.8' | Columbia | Bronfman et al (2000b – table 2) |
| | 328$^o$ | 8.8' | Columbia | Bronfman (2008 - sect. 4) |
| | 330$^o$ | 8.8' | Columbia | García et al (2014 – table 3) |
| | 330$^o$ | 8.8' | Columbia | Grabelsky et al (1987 – sect. 3.1.2) |
| | 328.4 ±0.9 | mean and r.m.s. | | s.d.m of 0.4$^o$, worth 37 pc at 7 kpc |
| Start of Perseus | 336$^o$ | 8.8' | Columbia | Bronfman et al (1989 – sect. 4) |
| | 336$^o$ | 8.8' | Columbia | Bronfman (2008 – sect. 4) |
| | 337$^o$ | 8.8' | Columbia | Alvarez et al (1990 – table 4) |
| | 337$^o$ | 8.8' | Columbia | Bronfman et al (2000a – sect. 3.4) |
| | 337$^o$ | 8.8' | Columbia | Bronfman et al (2000b - table 2) |
| | 337$^o$ | 8.8' | Columbia | Dame & Thaddeus (2008 – sect. 1) |
| | 338$^o$ | 8.8' | Columbia | García et al (2014 – table 3) |
| | 336.9 ±0.7 | mean and r.m.s. | | s.d.m of 0.3°, worth 37 pc at 8 kpc |
| Near 3-kpc arm | 026° | 7.5' | Columbia | Cohen et al (1980 – fig.3) |
| | 026° | 7.5' | Columbia | Dame et al (1986 – fig.9) |
| | 026° | | mean | |
| Scutum | 031° | 8.4' | CfA | Dame & Thaddeus (2011 – fig.4) |
| | 033° | 1.1' | NRAO | Sanders et al (1985 – fig.5b) |
| | 034° | 7.5' | Columbia | Cohen et al (1980 – fig.3) |
| | 035° | 1.0' | Texas | Chiar et al (1994 – Sect. 3) |
| | 033.2 ±1.7 | mean and r.m.s. | | s.d.m of 0.8°, worth 70 pc at 5 kpc |
| Sagittarius | 051° | 7.5' | Columbia | Cohen et al (1980 – fig.3) |
| | 051° | 7.5' | Columbia | Dame et al (1986 – fig.9) |
| | 051° | | mean | |

**Table 4 – Published HII complex results**

| Arm name | Tangent longitude | Range | Reference |
|---|---|---|---|
| Carina | 284° | optical | Russeil (2003 – table 6) |
| | 284° | radio | Downes et al (1980 – fig.4) |
| | 284° | mean | |
| Crux-Centaurus | 309° | optical | Russeil (2003 – table 6) |
| | 309° | radio | Downes et al (1980 – fig.4) |
| | 309° | mean | |
| Norma | 323° | optical | Russeil (2003 – table 6) |
| | 328° | radio | Downes et al (1980 – fig.4) |
| | 325.5 ±3.5° | mean and r.m.s. | |
| Near 3-kpc arm | 025° | optical | Russeil (2003 – table 6) |
| | 025° | radio | Sanders et al (1985 – fig. 5a) |
| | 025° | mean | |
| Scutum | 032° | optical | Russeil (2003 – table 6) |
| | 031° | radio | Downes et al (1980 – fig.4) |
| | 031.5° ±0.7° | mean and r.m.s. | |
| Sagittarius | 056° | optical | Russeil (2003 – table 6) |
| | 051° | optical | Russeil et al (2007 – fig.4) |
| | 046° | radio | Downes et al (1980 – fig.4) |
| | 051.0° ±5.0° | mean and r.m.s. | |

**Table 5 – Published maser results**

| Arm name | Tangent longitude | Range | Reference |
|---|---|---|---|
| Norma | 331.5° | methanol | Caswell et al (2011 – Sect. 4.6.2) |
| | 331.5° | mean | |
| Start of Perseus | 338° | methanol | Green et al (2011 – sect. 3.3.1) |
| | 338° | methanol | Green et al (2012 – Sect.2) |
| | 338.0° | mean | |
| Scutum | 026° | methanol | Green et al (2011 – sect. 3.3.1) |
| | 026° | methanol | Green et al (2012 – Sect.2) |
| | 030° | methanol, water | Reid et al (2014 – fig.1) |
| | 031° | methanol, water | Sato et al (2014 – fig3) |
| | 028.2° ±2.6° | mean and r.m.s. | |
| Sagittarius | 049.6° | methanol | Pandian & Goldsmith (2007 – sect.4) |
| | 050° | methanol, water | Reid et al (2014 – fig.1) |
| | 051° | methanol, water | Wu et al (2014 – sect. 4.2) |
| | 050.2° ±0.7° | mean and r.m.s. | |